\documentstyle[12pt,amssymb,amsmath]{article}
\newtheorem{theorem}{Theorem}[section]
\newtheorem{lemma}{Lemma}[section]

\newcommand{\p}{\partial}
\newcommand{\la}{\lambda}
\newcommand{\La}{\Lambda}

\newcommand{\s}{\sum_{i=1}^N}
\newcommand{\Bk}{(L^k)_{\geq 0}}
\newcommand{\Bn}{(L^n)_{\geq 0}}
\newcommand{\Qk}{(L^k)_{\geq 1}}
\newcommand{\Qn}{(L^n)_{\geq 1}}
\newcommand{\qn}{(\partial(L^n)_{\geq 1}\partial^{-1})^*}
\begin{document}

\title{ {\bf B\"{a}cklund
transformations for the KP and mKP hierarchies with
self-consistent sources}}
\author{ {\bf  Ting Xiao
    \hspace{1cm} Yunbo Zeng\dag } \\
    {\small {\it
    Department of Mathematical Sciences, Tsinghua University,
    Beijing 100084,People's Republic of China}} \\
    {\small {\it \dag
     Email: yzeng@math.tsinghua.edu.cn}}}

\date{}
\maketitle
\renewcommand{\theequation}{\arabic{section}.\arabic{equation}}

\begin{abstract}
Using gauge transformations for the corresponding generating
pseudo-differential operators $L^n$ in terms of eigenfunctions and
adjoint eigenfunctions, we construct several types of
auto-B\"{a}cklund transformations for the KP hierarchy with
self-consistent sources (KPHSCS) and mKP hierarchy with
self-consistent sources (mKPHSCS) respectively. The B\"{a}cklund
transformations from the KPHSCS to mKPHSCS are also constructed in
this way.
\end{abstract}

\hskip\parindent


\section{Introduction}
\setcounter{equation}{0} \hskip\parindent

Soliton equations with self-consistent sources (SESCSs) are
important models in many fields of physics, such as hydrodynamics,
solid state physics, plasma physics, etc. [5-13]. For example, the
KP equation with self-consistent sources (KPESCS) describes the
interaction of a long wave with a short-wave packet propagating on
the x,y plane at an angle to each other (see \cite{Mel'nikov87}
and the references therein). Until now, several methods have been
developed in solving the soliton hierarchies with self-consistent
sources (SHSCS) in 1+1 dimensions, such as the inverse scattering
transform(IST), Darboux transformation(DT) and so on
(see[5-7,9-19]). Extension to 2+1 dimensions of the established
methods investigated for SHSCSs in 1+1 dimensions is a subject of
current research. Recently, by treating the 2+1 dimensional
constrained soliton hierarchies \cite{Oevel933,Cheng95} as the
stationary ones of the corresponding hierarchies with
self-consistent sources, we develop a systematical way to find
integrable (2+1)-dimensional soliton hierarchies with
self-consistent sources and their Lax representations. For
example, in \cite{XiaoTing20041} and \cite{XiaoTing20042}, the
integrable KP hierarchy with self-consistent sources (KPHSCS) and
mKP hierarchy with self-consistent sources (mKPHSCS) together with
their Lax representations are obtained, and the generalized binary
Darboux transformations with arbitrary functions in time $t$ for
the KP equation with self-consistent sources (KPESCS) and mKP
equation with self-consistent sources (mKEHSCS) are constructed
to obtain some interesting solutions.\\

B\"{a}cklund transformations generated via gauge invariance have
been shown to be a powerful tool to investigate the soliton
hierarchies in the last three decades. The 2+1 dimensional
B\"{a}cklund transformations employed were constructed by the
so-called Dressing Method based on appropriate gauge
transformation \cite{Boiti85}. This dressing approach had been
earlier used to construct auto-B\"{a}cklund transformations for
the KP, two-dimensional three wave and Davey-Stewartson equations
in turn \cite{Matveev79,Levi81}. Konopelchenko, Oevel et al
extended this method to investigate some important 2+1 dimensional
integrable hierarchies constructed in the framework of Sato
theory, such as the KP hierarchy, the mKP hierarchy and the Dym
hierarchy [23-33]. Via gauge transformations utilizing
eigenfunctions and adjoint eigenfunctions, auto-B\"{a}cklund
transformations and B\"{a}cklund transformations between these
hierarchies are constructed which reveal the intimate connections
between them. Also, gauge transformations are applicable to
investigate the constrained KP and constrained mKP hierarchies
\cite{Oevel931,Oevel933,Shaw971,Shaw972}. These B\"{a}cklund
transformations map the bi-Hamiltonian structure of the
constrained KP hierarchy
 to that of the constrained mKP hierarchy 
\cite{Oevel933,Shaw971,Shaw972}.\\

From the Darboux transformations for the KP equation
\cite{Matveev91} and mKP equation \cite{Estevez}, the
auto-B\"{a}cklund transformations for the first equation in the
KPHSCS and mKPHSCS, i.e., for the KPESCS and mKPESCS have been
constructed respectively in \cite{XiaoTing20041} and
\cite{XiaoTing20042}. But the auto-B\"{a}cklund transformations
for these two hierarchies with self-consistent sources and the
relation between them still remain unknown. In this paper, we are
devoted to finding the gauge invariance of the KPHSCS and mKPHSCS
and the relation between them. Since in our approach we regard the
constrained hierarchy as the stationary hierarchy of the
corresponding hierarchy with self-consistent sources, the known
information about the constrained hierarchy may be suggestive for
us to obtain the information about the corresponding hierarchy
with self-consistent sources. The constrained KP hierarchy (cKPH)
studied in \cite{Oevel931,Shaw971,Shaw972} may be treated as the
stationary hierarchy of the KPHSCS, so it is straightforward for
us to generalize the auto-B\"{a}cklund transformations for the
cKPH studied in \cite{Oevel931,Shaw972} to those for the KPHSCS.
Though the constrained mKP hierarchy (cmKPH) considered in
\cite{Oevel933,Shaw971,Shaw972} is different from the stationary
hierarchy of the mKPHSCS in our case, the idea of constructing
B\"{a}cklund transformations for it is still helpful. In this
paper, utilizing the associated eigenfunctions and adjoint
eigenfunctions, we construct several types of auto-B\"{a}cklund
transformations for the KPHSCS and mKPHSCS respectively. Their
compositions and iteration are formulated. The B\"{a}cklund
transformations between them are also considered. As pointed out
in the paper, the results obtained here will recover some known
results for some degenerate cases obtained in
\cite{Oevel931,Shaw971,Shaw972}.\\

The paper is organized as follows. First, we briefly review some
notations and useful identities of the pseudo-differential
operators (PDO) in Section 2 and give some general relations for
PDOs under various gauge transformations in Section 3. In the
framework of Sato theory, the KPHSCS and mKPHSCS and their
conjugate Lax pairs are introduced in Section 4. Basing on Darboux
covariance of the Lax pairs, we construct several types of
auto-B\"{a}cklund transformations for the KPHSCS and mKPHSCS
respectively via gauge transformations in terms of the associated
eigenfunctions and adjoint eigenfunctions in Section 5 and Section
6. The B\"{a}cklund transformations from the KPHSCS to mKPHSCS are
also considered in
Section 7.

\section{Some notations and identities about the PDO}
\setcounter{equation}{0} \hskip\parindent We will discuss the
KPHSCS and mKPHSCS in the framework of Sato theory. First, we will
give some basic notations and identities about the PDO which will
be used in our following discussions. More details about the PDO
will be referred to [1-4].
\\
For a PDO of the form
$$\La=\sum_{i<\infty} u_i \p^i,$$
where $\p=\p_x$, $u_i\in g_0$ and $g_0$ is a differential algebra,
we have the following notations
$$(\La)_{\geq k}=\sum_{i\geq k} u_i \p^i,\ \ \ (\La)_{< k}=\sum_{i<k}u_i\p^i,\ \ \ (\La)_k=u_k\p^k,$$
$$res(\La)=u_{-1},\ \ \ (\La)^*=\sum_{i<\infty}(-1)^i\p^iu_i.$$
For a given function $f$, $\La f$ denotes the composition of $\La$
with the multiplication operator $f$ while $\La(f)$ denotes the
action of the differential part of $\La$ on $f$, i.e.,
$$\La(f)=(\La f)_0=[(\La)_{\geq 0} f]_0=\sum_{i\geq 0} u_i f^{(i)}.$$
Some identities of the PDO are also useful and we list them below
$$(\La^*)_0=res(\p^{-1}\La),\ \ \ (\La)_0=res(\La\p^{-1}),\ \ \ (\La\p^{-1})_{<0}=(\La)_0\p^{-1}+(\La)_{<0}\p^{-1},$$
$$[(\La)_{\geq 0}]^*=[\La^*]_{\geq 0},\ \ \ (\p^{-1}\La)_{<0}=\p^{-1}(\La^*)_0+\p^{-1}(\La)_{<0}.$$

\section{Gauge transformations}
\setcounter{equation}{0} \hskip\parindent Here we will give some
results about gauge transformations for the PDOs from which the
B\"{a}cklund transformations 
for the KPHSCS and mKPHSCS originate.
\\
\begin{lemma}
For arbitrary PDO $A$, functions $f$ and $g$, the following identities hold: \\
1)$(f^{-1}Af)_{\geq 1}=f^{-1}(A)_{\geq 0}f-f^{-1}(A)_{\geq 0}(f)$,\\
2)$(f\p f^{-1}Af\p^{-1}f^{-1})_{\geq 0}=f\p f^{-1}(A)_{\geq
0}f\p^{-1}f^{-1}-f(f^{-1}(A)_{\geq 0}(f))_x\p^{-1}f^{-1}$,\\
3)$(f^{-1}Af)_{\geq 1}=f^{-1}(A)_{\geq 1}f-f^{-1}(A)_{\geq 1}(f)$,\\
4)$(f_x^{-1}\p A \p^{-1}f_x)_{\geq 1}=f_x^{-1}\p(A)_{\geq
1}\p^{-1}f_x-f_x^{-1}((A)_{\geq 1}(f))_x$,\\
5)$(\p^{-1}gAg^{-1}\p)_{\geq 1}=\p^{-1}g(A)_{\geq
0}g^{-1}\p-\p^{-1}g^{-1}(A)_{\geq 0}^*(g)\p$,\\
6)$(g^{-1}\p^{-1}gAg^{-1}\p g)_{\geq 0}=g^{-1}\p^{-1}g(A)_{\geq
0}g^{-1}\p g+g^{-1}\p^{-1}g[g^{-1}(A_{\geq 0})^*(g)]_x$,\\
7)$(\p^{-1}g_xAg_x^{-1}\p)_{\geq 1}=\p^{-1}g_x(A)_{\geq
1}g_x^{-1}\p-\p^{-1}g_x^{-1}[(A)_{\geq 1}]^*(g_x)\p$,\\
8)$(\p^{-1}g\p A \p^{-1}g^{-1}\p)_{\geq 1}=\p^{-1}g\p(A)_{\geq
1}\p^{-1}g^{-1}\p-\p^{-1}g^{-1}[\p(A)_{\geq 1}\p^{-1}]^*(g)\p$.
\end{lemma}
The results of $1)-4)$ are given by Oevel and Rogers in
\cite{Oevel932} and the remaining can be proved directly.

\begin{lemma}
Let $L$ be an arbitrary PDO and $f(x,t_q)\neq 0$, $g(x,t_q)$ be
arbitrary functions, the following identities hold:\\

1)If $\tilde{L}=f^{-1}Lf$, $\tilde{f}=f^{-1}g$ and
$\tilde{g}=D^{-1}(fg)$, where $D^{-1}$ defines the integral
operation $D^{-1}(f)=\int^x f(\xi)d \xi$,
then\\
$(\tilde{L}^q)_{\geq 1}=[f^{-1}(L^q)_{\geq 0}f]_{\geq
1}$,\\
$\tilde{f}_{t_q}-(\tilde{L}^q)_{\geq
1}(\tilde{f})=-f^{-2}g[f_{t_q}-(L^q)_{\geq
0}(f)]+f^{-1}[g_{t_q}-(L^q)_{\geq 0}(g)]$,\\
$\tilde{g}_{t_q}+[\p(\tilde{L}^q)_{\geq
1}\p^{-1}]^*(\tilde{g})=\tilde{g}_{t_q}+D^{-1}([(\tilde{L}^q)_{\geq
1}]^*(\tilde {g}_x))=D^{-1}[f(g_{t_q}+(L^q)^*_{\geq
0}(g))]+D^{-1}[g(f_{t_q}-(L^q)_{\geq 0}(f))]$,\\

2)If $\tilde{L}=f\p f^{-1}Lf\p^{-1}f^{-1}$,
$\tilde{f}=f(f^{-1}g)_x$ and $\tilde{g}=f^{-1}D^{-1}(fg)$,
then\\
$(\tilde{L}^q)_{\geq 0}=[f\p f^{-1}(L^q)_{\geq
0}f\p^{-1}f^{-1}]_{\geq
0}$,\\
$\tilde{f}_{t_q}-(\tilde{L}^q)_{\geq
0}(\tilde{f})=-g[f^{-1}(f_{t_q}-(L^q)_{\geq
0}(f))]_x+f[f^{-1}(g_{t_q}-(L^q)_{\geq 0}(g))]_x$,\\
$\tilde{g}_{t_q}+(\tilde{L}^q)_{\geq 0}^*(\tilde{g})
=-f^{-2}D^{-1}(fg)[f_{t_q}-(L^q)_{\geq
0}(f)]+f^{-1}D^{-1}[f(g_{t_q}+(L^q)^*_{\geq 0}(g))]+f^{-1}D^{-1}[g(f_{t_q}-(L^q)_{\geq 0}(f))]$,\\

3)If $\tilde{L}=f^{-1}Lf$, $\tilde{f}=f^{-1}g$ and
$\tilde{g}=D^{-1}(fg_x)$,
then\\
$(\tilde{L}^q)_{\geq 1}=[f^{-1}(L^q)_{\geq 1}f]_{\geq
1}$,\\
$\tilde{f}_{t_q}-(\tilde{L}^q)_{\geq
1}(\tilde{f})=-f^{-2}g[f_{t_q}-(L^q)_{\geq
1}(f)]+f^{-1}[g_{t_q}-(L^q)_{\geq 1}(g)]$,\\
$\tilde{g}_{t_q}+[\p(\tilde{L}^q)_{\geq
1}\p^{-1}]^*(\tilde{g})=
D^{-1}[f(g_{t_q}+(\p(L^q)_{\geq
1}\p^{-1})^*(g))]+D^{-1}[g_x(f_{t_q}-(L^q)_{\geq 1}(f))]$,\\

4)If $\tilde{L}=f_x^{-1}\p L\p^{-1}f_x$, $\tilde{f}=f_x^{-1}g_x$
and $\tilde{g}=D^{-1}(f_xg)$,
then\\
$(\tilde{L}^q)_{\geq 1}=[f_x^{-1}\p (L^q)_{\geq 1}\p^{-1}f_x]_{\geq 1}$,\\
$\tilde{f}_{t_q}-(\tilde{L}^q)_{\geq
1}(\tilde{f})=-f_x^{-2}g_x[f_{t_q}-(L^q)_{\geq
1}(f)]_x+f_x^{-1}[g_{t_q}-(L^q)_{\geq 1}(g)]_x$,\\
$\tilde{g}_{t_q}+[\p(\tilde{L}^q)_{\geq 1}\p^{-1}]^*(\tilde{g})
=D^{-1}[f_x(g_{t_q}+(\p(L^q)_{\geq
1}\p^{-1})^*(g))]+D^{-1}[g(f_{t_q}-(L^q)_{\geq 1}(f))_x]$.\\
\end{lemma}
Parts of the above results (for $\tilde{f}$) are given by Oevel
and Rogers in \cite{Oevel932}, the remaining for $\tilde{g}$ can
be proved directly following Lemma 3.1.

\begin{lemma}
Let $L$ be an arbitrary PDO and $f(x,t_q)$, $g(x,t_q)\neq 0$ be
arbitrary functions, the following identities hold:\\

1)If $\tilde{L}=\p^{-1}gLg^{-1}\p$, $\tilde{f}=D^{-1}(fg)$ and
$\tilde{g}=fg^{-1}$,
then\\
$(\tilde{L}^q)_{\geq 1}=[\p^{-1}g(L^q)_{\geq 0}g^{-1}\p]_{\geq
1}$,\\
$\tilde{f}_{t_q}-(\tilde{L}^q)_{\geq
1}(\tilde{f})=D^{-1}[g(f_{t_q}-(L^q)_{\geq
0}(f))]+D^{-1}[f(g_{t_q}+(L^q)_{\geq 0}^*(g))]$,\\
$\tilde{g}_{t_q}+[\p(\tilde{L}^q)_{\geq
1}\p^{-1}]^*(\tilde{g})=g^{-1}[f_{t_q}+(L^q)_{\geq 0}^*(f)]-g^{-2}f[g_{t_q}+(L^q)^*_{\geq 0}(g)]$,\\

2)If $\tilde{L}=g^{-1}\p^{-1} gLg^{-1}\p g$,
$\tilde{f}=g^{-1}D^{-1}(fg)$ and $\tilde{g}=g(g^{-1}f)_x$,
then\\
$(\tilde{L}^q)_{\geq
0}=[g^{-1}\p^{-1} g(L^q)_{\geq 0}g^{-1}\p g]_{\geq 0}$,\\
$\tilde{f}_{t_q}-(\tilde{L}^q)_{\geq
0}(\tilde{f})=-g^{-2}D^{-1}(fg)[g_{t_q}+(L^q)_{\geq
0}^*(g)]+g^{-1}D^{-1}[g(f_{t_q}-(L^q)_{\geq 0}(f))]+g^{-1}D^{-1}[f(g_{t_q}+(L^q)_{\geq 0}^*(g))]$,\\
$\tilde{g}_{t_q}+(\tilde{L}^q)_{\geq 0}^*(\tilde{g})
=g[g^{-1}(f_{t_q}+(L^q)^*_{\geq 0}(f))]_x-f[g^{-1}(g_{t_q}+(L^q)^*_{\geq 0}(g))]_x$,\\

3)If $\tilde{L}=\p^{-1}g_xLg_x^{-1}\p$, $\tilde{f}=D^{-1}(fg_x)$
and $\tilde{g}=f_xg_x^{-1}$,
then\\
$(\tilde{L}^q)_{\geq 1}=[\p^{-1}g_x(L^q)_{\geq 1}g_x^{-1}\p]_{\geq
1}$,\\
$\tilde{f}_{t_q}-(\tilde{L}^q)_{\geq 1}(\tilde{f})=
D^{-1}[f(g_{t_q}+(\p(L^q)_{\geq 1}\p^{-1})^*(g))_x]+D^{-1}[g_x(f_{t_q}-(L^q)_{\geq 1}(f))]$,\\
$\tilde{g}_{t_q}+[\p(\tilde{L}^q)_{\geq 1}\p^{-1}]^*(\tilde{g})
=g_x^{-1}[f_{t_q}+(\p(L^q)_{\geq
1}\p^{-1})^*(f)]_x-g_x^{-2}f_x[g_{t_q}+(\p(L^q)_{\geq 1}\p^{-1})^*(g)]_x$,\\

4)If $\tilde{L}=\p^{-1}g\p L\p^{-1}g^{-1}\p$,
$\tilde{f}=D^{-1}(f_xg)$ and $\tilde{g}=fg^{-1}$,
then\\
$(\tilde{L}^q)_{\geq 1}=[\p^{-1}g\p (L^q)_{\geq
1}\p^{-1}g^{-1}\p]_{\geq 1}$,\\
$\tilde{f}_{t_q}-(\tilde{L}^q)_{\geq
1}(\tilde{f})=D^{-1}[f_x(g_{t_q}+(\p(L^q)_{\geq 1}\p^{-1})^*(g))]+D^{-1}[g(f_{t_q}-(L^q)_{\geq 1}(f))_x]$,\\
$\tilde{g}_{t_q}+[\p(\tilde{L}^q)_{\geq 1}\p^{-1}]^*(\tilde{g})
=g^{-1}[f_{t_q}+(\p(L^q)_{\geq
1}\p^{-1})^*(f)]-g^{-2}f[g_{t_q}+(\p(L^q)_{\geq 1}\p^{-1})^*(g)]$.\\
\end{lemma}
These results can be proved directly following Lemma 3.1.

\section{The KPHSCS and mKPHSCS}
\setcounter{equation}{0} \hskip\parindent 1.Consider the PDO of
the form
\begin{equation}
\label{1}
    L=L_{KP}=\partial+(u(t)/2)\partial^{-1}+u_1(t)\partial^{-2}+...,\
    \ t=(t_1=x,t_2,\cdots),
\end{equation}
which satisfies the constraint
\begin{equation}
\label{2} L^n=(L^n)_{\geq 0}+\s
q_i(t)\partial^{-1}r_i(t)=\p^n+\omega_{n-2}(t)\p^{n-2}+\cdots+\omega_0(t)+\s
q_i(t)\partial^{-1}r_i(t),\ \ \omega_{n-2}=\frac{n}{2}u,
\end{equation}
with $q_i(t), r_i(t)$ satisfy
$$(q_i)_{t_k}=(L^k)_{\geq 0}(q_i),
\ \ (r_i)_{t_k}=-(L^k)_{\geq 0}^*(r_i),\ \ k, n \in \mathbb{N},$$
where
$$(L^k)_{\geq 0}=[(L^n)^{\frac{k}{n}}]_{\geq 0}=[((L^n)_{\geq 0}+\s q_i\p^{-1}r_i)^{\frac{k}{n}}]_{\geq 0}.$$
Starting from the PDO $L^n$ (\ref{2}) and requiring $k<n$, the
KPHSCS is defined by the following
Lax representation \cite{XiaoTing20041}\\
\begin{subequations}
\label{3}
\begin{equation}
\label{31} ((L^k)_{\geq 0})_{t_n}-(L^n)_{t_k}+[(L^k)_{\geq
0},L^n]=0,
\end{equation}
\begin{equation}
\label{32} (q_i)_{t_k}=(L^k)_{\geq 0}(q_i),
\end{equation}
\begin{equation}
\label{33} (r_i)_{t_k}=-(L^k)_{\geq 0}^*(r_i),\ \ i=1,...,N.
\end{equation}
\end{subequations}
Under (\ref{32}) and (\ref{33}), (\ref{31}) will be obtained by
the compatibility condition of either
\begin{subequations}
\label{4}
\begin{equation}
\label{41}
     \psi_{t_k} =\Bk(\psi),
\end{equation}
\begin{equation}
\label{42}
    \psi_{t_n} = \Bn(\psi)+\s q_iD^{-1} (r_i\psi),
\end{equation}
\end{subequations}
or
\begin{subequations}
\label{5}
\begin{equation}
\label{51}
     \bar{\psi}_{t_k} =-\Bk^*(\bar{\psi}),
\end{equation}
\begin{equation}
\label{52}
    \bar{\psi}_{t_n} = -\Bn^*(\bar{\psi})+\s r_iD^{-1}
    (q_i\bar{\psi}).
\end{equation}
\end{subequations}
$\psi$ and $\bar{\psi}$ will be called the eigenfunction and
adjoint eigenfunction of the
KPHSCS (\ref{3}) respectively.\\
(\ref{3}) will be often written as equations of the fields $u$,
$q_i$ and $r_i$ when the auxiliary fields
$\omega_{n-3}$,$\cdots$,$\omega_0$ are eliminated. For example,
when $k=2$,$n=3$, we will get the KPESCS as
\cite{Mel'nikov87,XiaoTing20041}
 \begin{subequations}
\label{5.1}
\begin{equation}
\label{5.11}
     [4u_t-6uu_x-u_{xxx}+8(\sum_{j=1}^{N}q_jr_j)_x]_x-3u_{yy} =0,
\end{equation}
\begin{equation}
\label{5.12}
    q_{j,y}=q_{j,xx}+uq_j,
\end{equation}
\begin{equation}
\label{5.13}
    r_{j,y}=-r_{j,xx}-ur_j, \ \ \ j=1,...,N.
\end{equation}
\end{subequations}
\\
2.Consider another PDO of the form
\begin{equation}
\label{6}
    L=L_{mKP}=\partial+v(t)+v_1(t)\partial^{-1}+...,
\end{equation}
which satisfies the constraint
\begin{equation}
\label{7} L^n=(L^n)_{\geq 1}+\sum_{i=1}^N
q_i(t)\partial^{-1}r_i(t)\p=\p
^n+\pi_{n-1}(t)\p^{n-1}+\cdots+\pi_1(t)\p+\sum_{i=1}^N
q_i(t)\partial^{-1}r_i(t)\p,\ \ \pi_{n-1}=nv,
\end{equation}
with $q_i(t), r_i(t)$ satisfy
$$(q_i)_{t_k}=(L^k)_{\geq 1}(q_i),
\ \ (r_i)_{t_k}=-(\p(L^k)_{\geq 1}\p^{-1})^*(r_i),\ \ k, n \in
\mathbb{N},$$ where
$$(L^k)_{\geq 1}=[(L^n)^{\frac{k}{n}}]_{\geq 1}=[((L^n)_{\geq 1}+\s q_i\p^{-1}r_i\p)^{\frac{k}{n}}]_{\geq 1}.$$
Starting from the PDO (\ref{7}) and requiring $k<n$, the mKPHSCS
is defined by the following Lax
representation \cite{XiaoTing20042}\\
\begin{subequations}
\label{8}
\begin{equation}
\label{81} ((L^k)_{\geq 1})_{t_n}-(L^n)_{t_k}+[(L^k)_{\geq
1},L^n]=0,
\end{equation}
\begin{equation}
\label{82} (q_i)_{t_k}=(L^k)_{\geq 1}(q_i),
\end{equation}
\begin{equation}
\label{83} (r_i)_{t_k}=-(\p(L^k)_{\geq 1}\p^{-1})^*(r_i),\ \
i=1,...,N.
\end{equation}
\end{subequations}
Under (\ref{82}) and (\ref{83}), (\ref{81}) will be obtained by
the compatibility condition of either
\begin{subequations}
\label{9}
\begin{equation}
\label{91}
     \phi_{t_k} =\Qk(\phi),
\end{equation}
\begin{equation}
\label{92}
    \phi_{t_n} = \Qn(\phi)+\s q_iD^{-1} (r_i\phi_x),
\end{equation}
\end{subequations}
or
\begin{subequations}
\label{10}
\begin{equation}
\label{101}
     \bar{\phi}_{t_k} =-(\p \Qk \p^{-1})^*(\bar{\phi}),
\end{equation}
\begin{equation}
\label{102}
    \bar{\phi}_{t_n} = -(\p \Qn \p^{-1})^*(\bar{\phi})-\s r_iD^{-1}
    (q_i\bar{\phi}_x).
\end{equation}
\end{subequations}
$\phi$ and $\bar{\phi}$ will be called the eigenfunction and
adjoint eigenfunction of the mKPHSCS (\ref{8}) respectively.\\
(\ref{8}) will be often written as equations of the fields $v$,
$q_i$ and $r_i$ when the auxiliary fields
$\pi_{n-2}$,$\cdots$,$\pi_1$ are eliminated. For example, When
$k=2$,$n=3$, we will get the mKPESCS as \cite{XiaoTing20042}
\begin{subequations}
\label{10.1}
\begin{equation}
\label{10.11}
    4v_t-v_{xxx}-3D^{-1}(v_{yy})-6D^{-1}(v_y)v_x+6v^2v_x+4\sum_{i=1}^{N}(q_ir_i)_x=0,
\end{equation}
\begin{equation}
\label{10.12}
    q_{i,y}=q_{i,xx}+2vq_{i,x},
\end{equation}
\begin{equation}
\label{10.13} r_{i,y}=-r_{i,xx}+2vr_{i,x},\ \ \ \ i=1,...,N.
\end{equation}
\end{subequations}


The identities shown in Section 3 lead in a natural way to
invariances and relations of the KPHSCS and mKPHSCS. We will
discuss them in the following sections.

\section{The auto-B\"{a}cklund transformations for the KPHSCS}
\setcounter{equation}{0} \hskip\parindent 1. Auto-B\"{a}cklund
transformation utilizing eigenfunction.
\begin{theorem}
Let $L^n$ of (\ref{2}) satisfy the KPHSCS (\ref{3}) and $\psi$ be
the corresponding
eigenfunction. The function $f\neq 0$ satisfies (\ref{4}).\\
Define
$$T_1[f]: L^n \stackrel{f}{\longmapsto}
\tilde{L}^n$$ by
\begin{subequations}
\label{12}
\begin{equation}
\label{121} \tilde{L}^n=(\chi_1L^n\chi_1^{-1})_{\geq 0}+\s
\tilde{q}_i\p^{-1}\tilde{r}_i,
\end{equation}
where
\begin{equation}
\label{122} \chi_1=f\p f^{-1},\ \ \tilde{q}_i=f(f^{-1}q_i)_x, \ \
\ \tilde{r}_i=-f^{-1}D^{-1}(r_if),\ \ i=1,...,N.
\end{equation}
\end{subequations}
Then $\tilde{L}^n$ will also satisfy the KPHSCS (\ref{3}). So,
$T_1[f]$ defines an auto-B\"{a}cklund transformation for the
KPHSCS (\ref{3}).
\end{theorem}
{\bf{Proof}}:It is straightforward to see $(\tilde{L}^n)_{\geq 0}$
has the same form as $(L^n)_{\geq 0}$, i.e.,
$$(\tilde{L}^n)_{\geq 0}=(\chi_1L^n\chi_1^{-1})_{\geq 0}=\p^n+\tilde{\omega}_{n-2}\p^{n-2}+\cdots+\tilde{\omega_0},\ \  \tilde{\omega}_{n+2}=\frac{n\tilde{u}}{2}=\frac{n}{2}(u+2({\mathrm{ln}} f)_{xx}).$$
Furthermore, we can find that for $k<n$,
\begin{equation} \nonumber
\begin{array}{lll}
(\tilde{L}^k)_{\geq 0}&=&[(\tilde{L}^n)^{\frac{k}{n}}]_{\geq
0}=[((\tilde{L}^n)_{\geq 0})^{\frac{k}{n}}]_{\geq
0}=[((\chi_1L^n\chi_1^{-1})_{\geq 0})^{\frac{k}{n}}]_{\geq
0}=[(\chi_1L^n\chi_1^{-1})^{\frac{k}{n}}]_{\geq 0}\\
&=&[\chi_1(L^n)^{\frac{k}{n}}\chi_1^{-1}]_{\geq
0}=[\chi_1(L^k)_{\geq 0}\chi_1^{-1}]_{\geq 0}.
\end{array}
\end{equation}
Define $\tilde{\psi}=f(f^{-1}\psi)_x$. From Lemma3.2 2), we can
easily see that $\tilde{q}_i$,$\tilde{r}_i$ and $\tilde{\psi}$
satisfy (\ref{32}),(\ref{33}) and (\ref{41}) w.r.t. $\tilde{L}^n$
respectively. Besides,
\begin{equation}
\nonumber
\begin{array}{lll}
\tilde{\psi}_{t_n}-(\tilde{L}^n)_{\geq 0}(\tilde{\psi})&=&-\psi[f^{-1}(f_{t_n}-(L^n)_{\geq 0}(f))]_x+f[f^{-1}(\psi_{t_n}-(L^n)_{\geq 0}(\psi))]_x\\
&=&-\psi[f^{-1}\s q_iD^{-1}(r_if)]_x+f[f^{-1}\s q_iD^{-1}(r_i\psi)]_x\\
&=&-\psi\s (f^{-1}q_i)_xD^{-1}(r_if)+f\s (f^{-1}q_i)_xD^{-1}(r_i\psi)\\
&=&-\s (f^{-1}q_i)_xf[f^{-1}\psi D^{-1}(r_if)-D^{-1}(r_iff^{-1}\psi)]\\
&=&\s f(f^{-1}q_i)_xD^{-1}[-f^{-1}D^{-1}(r_if)f(f^{-1}\psi)_x] \\
&=&\s \tilde{q}_iD^{-1}(\tilde{r}_i\tilde{\psi}),
\end{array}
\end{equation}
so $\tilde{\psi}$ satisfies (\ref{42}) w.r.t.
$\tilde{L}^n$.\\
This completes the proof.\\
2. Auto-B\"{a}cklund transformation utilizing adjoint
eigenfunction.
\begin{theorem}
Let 
$L^n$ of (\ref{2}) satisfy the KPHSCS (\ref{3}) and $\psi$ be the
corresponding
eigenfunction. The function $g\neq 0$ satisfies (\ref{5}).\\
Define
$$T_2[g]: L^n \stackrel{g}{\longmapsto}
\tilde{L}^n$$ by
\begin{subequations}
\label{12}
\begin{equation}
\label{121} \tilde{L}^n=(\chi_2L^n\chi_2^{-1})_{\geq 0}+\s
\tilde{q}_i\p^{-1}\tilde{r}_i,
\end{equation}
where
\begin{equation}
\label{122} \chi_2=g^{-1}\p^{-1}g,\ \
\tilde{q_i}=g^{-1}D^{-1}(gq_i), \ \ \tilde{r_i}=-g(r_ig^{-1})_x,\
\ i=1,...,N.
\end{equation}
\end{subequations}
Then $\tilde{L}^n$ will also satisfy the KPHSCS (\ref{3}). So,
$T_2[g]$ defines an auto-B\"{a}cklund transformation for the
KPHSCS (\ref{3}).
\end{theorem}
{\bf{Proof}}: It is straightforward to see $(\tilde{L}^n)_{\geq
0}$ has the same form as $(L^n)_{\geq 0}$, i.e.,
$$(\tilde{L}^n)_{\geq 0}=(\chi_2L^n\chi_2^{-1})_{\geq 0}=\p^n+\tilde{\omega}_{n-2}\p^{n-2}+\cdots+\tilde{\omega_0},\ \  \tilde{\omega}_{n+2}=\frac{n\tilde{u}}{2}=\frac{n}{2}(u+2({\mathrm{ln}} g)_{xx}).$$
Furthermore, we can find that for $k<n$,
\begin{equation}
\nonumber
\begin{array}{lll}
(\tilde{L}^k)_{\geq 0}&=&[(\tilde{L}^n)^{\frac{k}{n}}]_{\geq
0}=[((\tilde{L}^n)_{\geq 0})^{\frac{k}{n}}]_{\geq
0}=[((\chi_2L^n\chi_2^{-1})_{\geq 0})^{\frac{k}{n}}]_{\geq
0}=[(\chi_2L^n\chi_2^{-1})^{\frac{k}{n}}]_{\geq 0}\\
&=&[\chi_2(L^n)^{\frac{k}{n}}\chi_2^{-1}]_{\geq
0}=[\chi_2(L^k)_{\geq 0}\chi_2^{-1}]_{\geq 0}.
\end{array}
\end{equation}
Define $\tilde{\psi}=g^{-1}D^{-1}(g\psi)$. From Lemma3.3 2), we
can easily see that $\tilde{q}_i$,$\tilde{r}_i$ and $\tilde{\psi}$
satisfy (\ref{32}),(\ref{33}) and (\ref{41}) w.r.t. $\tilde{L}^n$
respectively. Besides,
\begin{equation}
\nonumber
\begin{array}{ll}
&\tilde{\psi}_{t_n}-(\tilde{L}^n)_{\geq 0}(\tilde{\psi})\\
=&-g^{-2}D^{-1}(g\psi)[g_{t_n}+(L^n)_{\geq 0}^*(g)]+g^{-1}D^{-1}[g(\psi_{t_n}-(L^n)_{\geq 0}(\psi))]+g^{-1}D^{-1}[\psi(g_{t_n}+(L^n)_{\geq 0}^*(g))]\\
=&-g^{-2}D^{-1}(g\psi)\s r_iD^{-1}(gq_i)+g^{-1}D^{-1}[g\s
q_iD^{-1}(r_i\psi)]+g^{-1}D^{-1}[\psi\s r_iD^{-1}(gq_i)]\\
=&-g^{-1}\s D^{-1}(gq_i)D^{-1}(r_i\psi)-g^{-2}\s
D^{-1}(g\psi)D^{-1}(gq_i)r_i\\
=&-\s g^{-1}
D^{-1}(gq_i)[-D^{-1}(r_ig^{-1}g\psi)+g^{-1}r_iD^{-1}(g\psi)]\\
=&-\s g^{-1} D^{-1}(gq_i)D^{-1}[(g^{-1}r_i)_xD^{-1}(g\psi)]\\
=&\s \tilde{q}_iD^{-1}(\tilde{r}_i\tilde{\psi}),
\end{array}
\end{equation}
so $\tilde{\psi}$ satisfies (\ref{42}) w.r.t.
$\tilde{L}^n$.\\
This completes the proof.\\

3. $T$: composition of $T_1$ and $T_2$.\\
Let $L^n$ of (\ref{2}) be a solution of the KPHSCS (\ref{3}),
$f\neq 0$ and $g\neq 0$ satisfy (\ref{4}) and (\ref{5})
respectively. Utilizing $f$, $T_1[f]$ transforms $L^n$ into
$\tilde{L}^n=(\tilde{L}^n)_{\geq 0}+\s
\tilde{q}_i\p^{-1}\tilde{r}_i$ which is a new solution of the
KPHSCS (\ref{3}). It is easy to prove that
$\tilde{g}=f^{-1}D^{-1}(fg)$ satisfies (\ref{5}) w.r.t.
$\tilde{L}^n$. So utilizing $\tilde{g}$ again, $\tilde{L}^n$ will
be transformed by $T_2[\tilde{g}]$ into
$\hat{L}^n=(\hat{L}^n)_{\geq
0}+\sum_{j=1}^N\hat{q}_j\partial^{-1}\hat{r}_j=\p^n+\hat{\omega}_{n-2}\p^{n-2}+\cdots+\hat{\omega_0}+\sum_{j=1}^N\hat{q}_j\partial^{-1}\hat{r}_j$
($\hat{\omega}_{n-2}=\frac{n}{2}\hat{u}$) which will satisfy the
KPHSCS (\ref{3}) again. We list the results below.
\begin{subequations}
\label{15}
\begin{equation}
\label{151}
\begin{array}{lll}
\hat{L}^n&\stackrel{\triangle}{=}&T[f,g](L^n)=(T_2[\tilde{g}]\circ
T_1[f])(L^n)=(\chi L^n\chi^{-1})_{\geq
0}+\sum_{j=1}^N\hat{q}_j\partial^{-1}\hat{r}_j,\\
&&\chi=\tilde{g}^{-1}\p^{-1}\tilde{g}f\p
f^{-1},(\tilde{g}=f^{-1}D^{-1}(fg)),\\
\end{array}
\end{equation}
\begin{equation}
\label{152}
\hat{u}=u+2[{\mathrm{ln}}(D^{-1}(fg))]_{xx},\\
\end{equation}
\begin{equation}
\label{153}
\hat{q_i}\stackrel{\triangle}{=}T[f,g](q_i)=(T_2[\tilde{g}]\circ
T_1[f])(q_i)=q_i-\frac{fD^{-1}(q_i g)}{D^{-1}(fg)}, \\
\end{equation}
\begin{equation}
\label{154}
\hat{r_i}\stackrel{\triangle}{=}T[f,g](q_i)=(T_2[\tilde{g}]\circ
T_1[f])(r_i)=r_i-\frac{gD^{-1}(r_i f)}{D^{-1}(fg)}. \\
\end{equation}
\end{subequations}

{\bf{Remark}}:\\

1.When $((L^k)_{\geq 0})_{t_n}=0$, (\ref{3}) will degenerate to
the constrained KP hierarchy. Correspondingly, $T_1[f]$ with $f$
satisfying (\ref{41}) and $T_2[g]$ with $g$ satisfying (\ref{51})
will give auto-B\"{a}cklund transformations for the constrained KP
hierarchy (be strictly, not the whole hierarchy because $k<n$).
Different from our case that the auto-B\"{a}cklund transformations
here are constructed directly for $L^n$ (\ref{2}), that
constructed in \cite{Oevel931} is for $L$ (\ref{1}). So in order
to preserve the constraint form of (\ref{2}), $f$ used in
\cite{Oevel931} has to satisfy an extra eigenvalue problem $\Bn
(f)+\s q_iD^{-1}
(r_if)=\la f$.\\

2.When $q_j=r_j=0$,$j=1,\cdots,N$, (\ref{3}) will degenerate to
the KP hierarchy. $T_1[f]$,$T_2[g]$ and $T[f,g]$ introduced above
will degenerate to the auto-B\"{a}cklund transformations for the
KP hierarchy based on gauge transformations which have been widely
studied in [24-31]($f$ and $g$ now only need to be eigenfunction
and adjoint eigenfunction of the KP hierarchy,
i.e., satisfy (\ref{4}) and (\ref{5}) respectively with $q_i=r_i=0$);\\

3.When $k=2$ and $n=3$, $T_1[f]$, $T_2[g]$ and $T[f,g]$ introduced
above will give rise to the auto-B\"{a}cklund transformations for
the KPESCS (\ref{5.1}) induced by the backward, forward and binary
Darboux transformations respectively \cite{XiaoTing20041}($f$ and
$g$ now only need to be eigenfunction and adjoint eigenfunction
of the KPESCS respectively).\\

4.({\bf{Iteration of $T$}})\\
Assuming $\psi_1,...,\psi_n$ are $n$ arbitrary solutions of
(\ref{4}) and $\bar{\psi}_1,...,\bar{\psi}_n$ are $n$ arbitrary
solutions of (\ref{5}), we define the following Wronskians (Gramm
Determinants):
\begin{equation}
\label{16}
\begin{array}{lll}
W_1(\psi_1,...,\psi_n;\bar{\psi}_1,...,\bar{\psi}_n)&=&det(U_{n\times
n}),\\W_2(\psi_1,...,\psi_n;\bar{\psi}_1,...,\bar{\psi}_{n-1})&=&det(V_{n\times
n}),\\W_3(\psi_1,...,\psi_{n-1};\bar{\psi}_1,...,\bar{\psi}_n)&=&det(X_{n\times
n}),
\end{array}
\end{equation}
where
\begin{subequations}
\label{16.1}
\begin{equation}
\label{16.11}
     U_{i,j}=D^{-1}(\bar{\psi}_i\psi_j),\ \ \  i,j=1,...,n,
\end{equation}
\begin{equation}
\label{16.12} V_{i,j}=U_{i,j},\ i=1,...,n-1,\ j=1,...,n;\ \
V_{n,j}=\psi_j,\ j=1,...,n,
\end{equation}
\begin{equation}
\label{16.13} X_{i,j}=D^{-1}(\bar{\psi}_j\psi_i),\ i=1,...,n-1,\
j=1,...,n;\ \ X_{n,j}=\bar{\psi}_j,\ j=1,...,n.
\end{equation}
\end{subequations}
We denote $u[n]$, $q_i[n]$ and $r_i[n]$ as the $n$ times iteration
of $T$ on $u$, $q_i$ and $r_i$ respectively, then the following
formulas hold.
\begin{subequations}
\label{16.2}
\begin{equation}
\label{16.21}
     u[n]=u+2\partial^2\mathrm{ln} W_1(f_1,...,
f_n;g_1,..., g_n),
\end{equation}
\begin{equation}
\label{16.22} q_i[n]=\frac {W_2(f_1,..., f_n,q_i;g_1,...,
g_n)}{W_1(f_1,..., f_n;g_1,..., g_n)},
\end{equation}
\begin{equation}
\label{16.23} r_i[n]=\frac {W_3(f_1,..., f_n;g_1,...,
g_n,r_i)}{W_1(f_1,..., f_n;g_1,..., g_n)},
\end{equation}
\end{subequations}
where $f_i$ and $g_i$, $i=1,\cdots,n$ are arbitrary solutions of
(\ref{4}) and (\ref{5}) respectively. \\
We can prove (\ref{16.2}) in the same way as we did in
\cite{XiaoTing20041} and we omit it here.

\section{The auto-B\"{a}cklund transformations for the mKPHSCS}
\setcounter{equation}{0} \hskip\parindent 1. Auto-B\"{a}cklund
transformations utilizing eigenfunction.
\begin{theorem}
Let $L^n$ of (\ref{7}) satisfy the mKPHSCS (\ref{8}) and $\phi$ is
the corresponding eigenfunction. The function $f\neq 0$ satisfies
(\ref{9}).Define
$$G_1[f]: L^n\stackrel{f}{\longmapsto}
\tilde{L}^n$$ by
\begin{equation}
\label{17} \tilde{L}^n=(f^{-1}L^nf)_{\geq
1}+\sum_{j=1}^N\tilde{q}_j\partial^{-1}\tilde{r}_j\p,\ \
\tilde{q}_j=f^{-1}q_j, \ \ \tilde{r}_j=D^{-1}(r_{j,x}f),\ \
j=1,...,N.
\end{equation}
Then $\tilde{L}^n$ will also satisfy the mKPHSCS (\ref{8}).
\end{theorem}
{\bf{Proof}}:It is straightforward to see $(\tilde{L}^n)_{\geq 1}$
has the same form as $(L^n)_{\geq 1}$, i.e.,
$$(\tilde{L}^n)_{\geq 1}=(f^{-1}L^nf)_{\geq
1}=\p^n+\tilde{\pi}_{n-1}\p^{n-1}+\cdots+\tilde{\pi}_1\p,\ \
\tilde{\pi}_{n-1}=n\tilde{v}=n(v+({\mathrm{ln}} f)_x).$$
Furthermore we can find that for $k<n$,
\begin{equation}
\nonumber
\begin{array}{lll}
(\tilde{L}^k)_{\geq 1}&=&[(\tilde{L}^n)^{\frac{k}{n}}]_{\geq
1}=[((\tilde{L}^n)_{\geq 1})^{\frac{k}{n}}]_{\geq 1}=[((f^{-1}L^n
f)_{\geq 1})^{\frac{k}{n}}]_{\geq
1}=[(f^{-1}L^nf)^{\frac{k}{n}}]_{\geq 1}\\
&=&[f^{-1}(L^n)^{\frac{k}{n}}f]_{\geq 1}=[f^{-1}(L^k)_{\geq
1}f]_{\geq 1}.
\end{array}
\end{equation}
Define $\tilde{\phi}=f^{-1}\phi$. From Lemma3.2 3), we can easily
see that $\tilde{q}_i$,$\tilde{r}_i$,$i=1,...,N$ and
$\tilde{\phi}$ satisfy (\ref{82}),(\ref{83}) and (\ref{91}) w.r.t.
$\tilde{L}^n$ respectively. Besides,
\begin{equation}
\nonumber
\begin{array}{lll}
\tilde{\phi}_{t_n}-(\tilde{L}^n)_{\geq 1}(\tilde{\phi})&=&-f^{-2}\phi[f_{t_n}-(L^n)_{\geq 1}(f)]+f^{-1}[\phi_{t_n}-(L^n)_{\geq 1}(\phi)]\\
&=&-f^{-2}\phi\s q_iD^{-1}(r_if_x)+f^{-1}[\s q_iD^{-1}(r_i\phi_x)]\\
&=&f^{-2}\phi\s q_iD^{-1}(r_{i,x}f)-f^{-1}\s q_iD^{-1}(r_{i,x}\phi)\\
&=&\sum_{i=1}^N \tilde{q}_iD^{-1}(\tilde{r}_i\tilde{\phi}_x).
\end{array}
\end{equation}
So $\tilde{\phi}$ satisfies (\ref{92}) w.r.t. $\tilde{L}^n$.
This completes the proof.\\

\begin{theorem}
Let 
$L^n$ of (\ref{7}) satisfy the mKPHSCS (\ref{8}) and $\phi$ be the
corresponding eigenfunction. The function $f\neq 0$ satisfies
(\ref{9}).Define
$$G_2[f]: L^n\stackrel{f}{\longmapsto}
\tilde{L}^n$$ by
\begin{equation}
\label{17} \tilde{L}^n=(f_x^{-1}\p L^n \p^{-1}f_x)_{\geq
1}+\sum_{j=1}^N\tilde{q}_j\partial^{-1}\tilde{r}_j\p,\ \
\tilde{q}_j=f_x^{-1}q_{j,x}, \ \ \tilde{r}_j=-D^{-1}(r_jf_x),\ \
j=1,...,N.
\end{equation}
Then $\tilde{L}^n$ will also satisfy the mKPHSCS (\ref{8}).
\end{theorem}
{\bf{Proof}}:It is straightforward to see $(\tilde{L}^n)_{\geq 1}$
has the same form as $(L^n)_{\geq 1}$, i.e.,
$$(\tilde{L}^n)_{\geq 1}=(f_x^{-1}\p L^n\p^{-1}f_x)_{\geq
1}=\p^n+\tilde{\pi}_{n-1}\p^{n-1}+\cdots+\tilde{\pi}_1\p,\ \
\tilde{\pi}_{n-1}=n\tilde{v}=n(v+({\mathrm{ln}} f_x)_x).$$
Furthermore we can find that for $k<n$,
\begin{equation}
\nonumber
\begin{array}{lll}
(\tilde{L}^k)_{\geq 1}&=&[(\tilde{L}^n)^{\frac{k}{n}}]_{\geq
1}=[((\tilde{L}^n)_{\geq 1})^{\frac{k}{n}}]_{\geq 1}=[((f_x^{-1}\p
L^n\p^{-1}f_x)_{\geq 1})^{\frac{k}{n}}]_{\geq
1}=[(f_x^{-1}\p L^n\p^{-1}f_x)^{\frac{k}{n}}]_{\geq 1}\\
&=&[f_x^{-1}\p (L^n)^{\frac{k}{n}}\p^{-1}f_x]_{\geq 1}=[f_x^{-1}\p
(L^k)_{\geq 1}\p^{-1}f_x]_{\geq 1}.
\end{array}
\end{equation}
Define $\tilde{\phi}=f_x^{-1}\phi_x$. From Lemma3.2 4), we can
easily see that $\tilde{q}_i$,$\tilde{r}_i$,$i=1,...,N$ and
$\tilde{\phi}$ satisfy (\ref{82}),(\ref{83}) and (\ref{91}) w.r.t.
$\tilde{L}^n$ respectively. Besides,
\begin{equation}
\nonumber
\begin{array}{lll}
\tilde{\phi}_{t_n}-(\tilde{L}^n)_{\geq 1}(\tilde{\phi})&=&-f_x^{-2}\phi_x[f_{t_n}-(L^n)_{\geq 1}(f)]_x+f_x^{-1}[\phi_{t_n}-(L^n)_{\geq 1}(\phi)]_x\\
&=&-f_x^{-2}\phi_x[\s q_iD^{-1}(r_if_x)]_x+f_x^{-1}[\s q_iD^{-1}(r_i\phi_x)]_x\\
&=&-f_x^{-2}\phi_x\s q_{i,x}D^{-1}(r_if_x)+f_x^{-1}\s q_{i,x}D^{-1}(r_i\phi_x)\\
&=&\sum_{i=1}^N \tilde{q}_iD^{-1}(\tilde{r}_i\tilde{\phi}_x),
\end{array}
\end{equation}
so $\tilde{\phi}$ satisfies (\ref{92}) w.r.t.
$\tilde{L}^n$.\\
This completes the proof.\\

2. Auto-B\"{a}cklund transformations utilizing adjoint
eigenfunction.
\begin{theorem}
Let $L^n$ of (\ref{7}) 
satisfy the mKPHSCS (\ref{8}) and $\phi$ be the
corresponding eigenfunction. The function $g$ ($g_x\neq 0$)
satisfies (\ref{10}).Define
$$G_3[g]: L^n\stackrel{g}{\longmapsto}
\tilde{L}^n$$ by
\begin{equation}
\label{17} \tilde{L}^n=(\p^{-1}g_x L^ng_x^{-1}\p)_{\geq
1}+\sum_{j=1}^N\tilde{q}_j\partial^{-1}\tilde{r}_j\p,\ \
\tilde{q}_j=-D^{-1}(g_xq_j), \ \ \tilde{r}_j=r_{j,x}g_x^{-1},\ \
j=1,...,N.
\end{equation}
Then $\tilde{L}^n$ will also satisfy the mKPHSCS (\ref{8}).
\end{theorem}
{\bf{Proof}}: It is straightforward to see $(\tilde{L}^n)_{\geq
1}$ has the same form as $(L^n)_{\geq 1}$, i.e.,
$$(\tilde{L}^n)_{\geq 1}=(\p^{-1}g_x L^ng_x^{-1}\p)_{\geq
1}=\p^n+\tilde{\pi}_{n-1}\p^{n-1}+\cdots+\tilde{\pi}_1\p,\ \
\tilde{\pi}_{n-1}=n\tilde{v}=n(v-({\mathrm{ln}} g_x)_x).$$
Furthermore we can find that for $k<n$,
\begin{equation}
\nonumber
\begin{array}{lll}
(\tilde{L}^k)_{\geq 1}&=&[(\tilde{L}^n)^{\frac{k}{n}}]_{\geq
1}=[((\tilde{L}^n)_{\geq 1})^{\frac{k}{n}}]_{\geq 1}=[((\p^{-1}g_x
L^ng_x^{-1}\p)_{\geq 1})^{\frac{k}{n}}]_{\geq
1}=[(\p^{-1}g_x L^ng_x^{-1}\p)^{\frac{k}{n}}]_{\geq 1}\\
&=&[\p^{-1}g_x (L^n)^{\frac{k}{n}}g_x^{-1}\p]_{\geq 1}=[\p^{-1}g_x
(L^k)_{\geq 1}g_x^{-1}\p]_{\geq 1}.
\end{array}
\end{equation}
Define $\tilde{\phi}=D^{-1}(g_x\phi)$. From Lemma3.3 3), we can
easily see that $\tilde{q}_i$,$\tilde{r}_i$,$i=1,...,N$ and
$\tilde{\phi}$ satisfy (\ref{82}),(\ref{83}) and (\ref{91}) w.r.t.
$\tilde{L}^n$ respectively. Besides,
\begin{equation}
\nonumber
\begin{array}{lll}
\tilde{\phi}_{t_n}-(\tilde{L}^n)_{\geq 1}(\tilde{\phi})&=&D^{-1}[g_x(\phi_{t_n}-(L^n)_{\geq 1}(\phi))]+D^{-1}[\phi(g_{t_n}+\qn(g))_x]\\
&=&D^{-1}[g_x(\s q_iD^{-1}(r_i\phi_x))]+D^{-1}[\phi(-\s r_iD^{-1}(q_ig_x))_x]\\
&=&D^{-1}[\s g_xq_i(r_i\phi-D^{-1}(r_{i,x}\phi))]-D^{-1}[\s \phi r_{i,x}D^{-1}(q_ig_x)+\phi r_ig_xq_i]\\
&=&-\s D^{-1}[D^{-1}(r_{i,x}\phi)q_ig_x+\phi
r_{i,x}D^{-1}(q_ig_x)]\\
&=&-\s D^{-1}(r_{i,x}\phi)D^{-1}(q_ig_x)\\
&=&\s \tilde{q}_iD^{-1}(\tilde{r}_i\tilde{\phi}_x),
\end{array}
\end{equation}
so $\tilde{\phi}$ satisfies (\ref{92}) w.r.t.
$\tilde{L}^n$.\\
This completes the proof.\\

\begin{theorem}
Let $L^n$ of (\ref{7})
satisfy the mKPHSCS (\ref{8}) and $\phi$ be the
corresponding eigenfunction. The function $g\neq 0$ satisfies
(\ref{10}).Define
$$G_4[g]: L^n\stackrel{g}{\longmapsto}
\tilde{L}^n$$ by
\begin{equation}
\label{17} \tilde{L}^n=(\p^{-1}g\p L^n\p^{-1}g^{-1}\p)_{\geq
1}+\sum_{j=1}^N\tilde{q}_j\partial^{-1}\tilde{r}_j\p,\ \
\tilde{q}_j=D^{-1}(gq_{j,x}), \ \ \tilde{r}_j=r_jg^{-1},\ \
j=1,...,N.
\end{equation}
Then $\tilde{L}^n$ will also satisfy the mKPHSCS (\ref{8}).
\end{theorem}
{\bf{Proof}}: It is straightforward to see $(\tilde{L}^n)_{\geq
1}$ has the same form as $(L^n)_{\geq 1}$, i.e.,
$$(\tilde{L}^n)_{\geq 1}=(\p^{-1}g\p
L^n\p^{-1}g^{-1}\p)_{\geq
1}=\p^n+\tilde{\pi}_{n-1}\p^{n-1}+\cdots+\tilde{\pi}_1\p,\ \
\tilde{\pi}_{n-1}=n\tilde{v}=n(v-({\mathrm{ln}} g)_x).$$
Furthermore we can find that for $k<n$,
\begin{equation}
\nonumber
\begin{array}{lll}
(\tilde{L}^k)_{\geq 1}&=&[(\tilde{L}^n)^{\frac{k}{n}}]_{\geq
1}=[((\tilde{L}^n)_{\geq 1})^{\frac{k}{n}}]_{\geq 1}=[((\p^{-1}g\p
L^n\p^{-1}g^{-1}\p)_{\geq 1})^{\frac{k}{n}}]_{\geq
1}=[(\p^{-1}g\p L^n\p^{-1}g^{-1}\p)^{\frac{k}{n}}]_{\geq 1}\\
&=&[\p^{-1}g\p (L^n)^{\frac{k}{n}}\p^{-1}g^{-1}\p]_{\geq
1}=[\p^{-1}g\p  (L^k)_{\geq 1}\p^{-1}g^{-1}\p]_{\geq 1}.
\end{array}
\end{equation}
Define $\tilde{\phi}=D^{-1}(g\phi_x)$. From Lemma3.3 4), we can
easily see that $\tilde{q}_i$,$\tilde{r}_i$,$i=1,...,N$ and
$\tilde{\phi}$ satisfy (\ref{82}),(\ref{83}) and (\ref{91}) w.r.t.
$\tilde{L}^n$ respectively. Besides,
\begin{equation}
\nonumber
\begin{array}{lll}
\tilde{\phi}_{t_n}-(\tilde{L}^n)_{\geq 1}(\tilde{\phi})&=&D^{-1}[g(\phi_{t_n}-(L^n)_{\geq 1}(\phi))_x]+D^{-1}[\phi_x(g_{t_n}+\qn(g))]\\
&=&D^{-1}[g(\s q_iD^{-1}(r_i\phi_x))_x]+D^{-1}[\phi_x(-\s r_iD^{-1}(q_ig_x))]\\
&=&D^{-1}[\s g(q_ir_i\phi_x+q_{i,x}D^{-1}(r_i\phi_x))]\\
&&-D^{-1}[\s \phi_x r_i(gq_i-D^{-1}(q_{i,x}g))]\\
&=&\s D^{-1}(gq_{i,x})D^{-1}(\phi_xr_i)\\
&=&\sum_{i=1}^n \tilde{q}_iD^{-1}(\tilde{r}_i\tilde{\phi}_x),
\end{array}
\end{equation}
so $\tilde{\phi}$ satisfies (\ref{92}) w.r.t.
$\tilde{L}^n$.\\
This completes the proof.\\

3.Compositions of $G_i$.\\

(1)$G_{12}$:composition of $G_1$ and $G_2$.\\
Let $L^n$ of (\ref{7}) satisfy the mKPHSCS (\ref{8}), $f_1$
($f_1\neq 0$ and $f_{1,x}\neq 0$) and $f_2=1$ satisfy (\ref{9}).
Utilizing $f_1$, $G_1[f_1]$ transforms $L^n$ into
$\tilde{L}^n=(\tilde{L}^n)_{\geq
1}+\sum_{j=1}^N\tilde{q}_j\partial^{-1}\tilde{r}_j\p$, a new
solution of the mKPHSCS (\ref{8}). It is not difficult to prove
that $\tilde{f_2}=f_1^{-1}f_2=f_1^{-1}$ satisfies (\ref{9}) w.r.t.
$\tilde{L}^n$. So utilizing $\tilde{f}_2$ again,
$G_2[\tilde{f}_2]$ transforms $\tilde{L}^n$ into another solution
$\hat{L}^n=(\hat{L}^n)_{\geq
1}+\sum_{j=1}^N\hat{q}_j\partial^{-1}\hat{r}_j\p$. 
We list the results below.
\begin{subequations}
\label{22}
\begin{equation}
\label{221}
\hat{L}^n\stackrel{\triangle}{=}G_{12}[f_1](L^n)=(G_2[\tilde{f}_2]\circ
G_1[f_1])(L^n)=(f_{1,x}^{-1}f_1^2\p
f_1^{-1}L^nf_1\p^{-1}f_1^{-2}f_{1,x})_{\geq 1}+\sum_{j=1}^N\hat{q}_j\partial^{-1}\hat{r}_j\p,\\
\end{equation}
\begin{equation}
\label{222}
\hat{q}_j\stackrel{\triangle}{=}G_{12}[f_1](q_j)=(G_2[\tilde{f}_2]\circ
G_1[f_1])(q_j)=f_{1,x}^{-1}(f_{1,x}q_j-f_1q_{j,x}), \\
\end{equation}
\begin{equation}
\label{223}
\hat{r}_j\stackrel{\triangle}{=}G_{12}[f_1](r_j)=(G_2[\tilde{f}_2]\circ
G_1[f_1])(r_j)=f_1^{-1}D^{-1}(r_jf_{1,x}), \ \ j=1,...,N.
\end{equation}
\end{subequations}

(2)$G_{34}$:composition of $G_3$ and $G_4$.\\
Let $L^n$ of (\ref{7}) 
satisfy the mKPHSCS (\ref{8}), $g_1$ ($g_1\neq 0$
and $g_{1,x}\neq 0$) and $g_2=1$ satisfy (\ref{10}). Utilizing
$g_1$, $G_3[g_1]$ transforms $L^n$ into
$\tilde{L}^n=(\tilde{L}^n)_{\geq
1}+\sum_{j=1}^N\tilde{q}_j\partial^{-1}\tilde{r}_j\p$, a new
solution of the mKPHSCS (\ref{8}). It is not difficult to prove
that $\tilde{g_2}=g_1^{-1}g_2=g_1^{-1}$ satisfies (\ref{10})
w.r.t. $\tilde{L}^n$. So utilizing $\tilde{g}_2$ again,
$G_4[\tilde{g}_2]$ transforms $\tilde{L}^n$ into
$\hat{L}^n=(\hat{L}^n)_{\geq
1}+\sum_{j=1}^N\hat{q}_j\partial^{-1}\hat{r}_j\p$. 
We list the results below.
\begin{subequations}
\label{23}
\begin{equation}
\label{231}
\begin{array}{lll}
\hat{L}^n&\stackrel{\triangle}{=}&G_{34}[g_1](L^n)=(G_4[\tilde{g}_2]\circ
G_3[g_1])(L^n)\\
&=&(\p^{-1}g_{1,x}g_1^{-2}\p^{-1}g_1\p
L^n\p^{-1}g_1^{-1}\p g_{1,x}^{-1}g_1^2\p)_{\geq 1}+\sum_{j=1}^N\tilde{q}_j\partial^{-1}\tilde{r}_j\p,\\
\end{array}
\end{equation}
\begin{equation}
\label{232}
\hat{q}_j\stackrel{\triangle}{=}G_{34}[g_1](q_j)=(G_4[\tilde{g}_2]\circ
G_3[g_1])(q_j)=g_1^{-1}D^{-1}(q_jg_{1,x}), \\
\end{equation}
\begin{equation}
\label{233}
\hat{r}_j\stackrel{\triangle}{=}G_{34}[g_1](q_j)=(G_4[\tilde{g}_2]\circ
G_3[g_1])(r_j)=g_{1,x}^{-1}(g_{1,x}r_j-g_1r_{j,x}), \ \ j=1,...,N.
\end{equation}
\end{subequations}

(3)$G$: composition of $G_{12}$ and $G_{34}$.\\
Let $L^n$ of (\ref{7}) satisfy the mKPHSCS (\ref{8}), $f$ ($f\neq
0$ and $f_x\neq 0$) and $g$ ($g\neq 0$ and $g_x\neq 0$) satisfy
(\ref{9}) and (\ref{10}) respectively. Utilizing $f$, $G_{12}[f]$
transforms $L^n$ into $\tilde{L}^n=(\tilde{L}^n)_{\geq
1}+\sum_{j=1}^N\tilde{q}_j\partial^{-1}\tilde{r}_j\p$. It is not
difficult to verify that
$\tilde{g}=G_{12}[f](g)=f^{-1}D^{-1}(gf_x)$ satisfies (\ref{10})
w.r.t. $\tilde{L}^n$. So utilizing $\tilde{g}$ again,
$G_{34}[\tilde{g}]$ transforms $\tilde{L}^n$ into another solution
of mKPHSCS (\ref{8}) $\hat{L}^n$. We list the results below.
\begin{subequations}
\label{24}
\begin{equation}
\label{241}
\begin{array}{lll}
\hat{L}^n&\stackrel{\triangle}{=}&G[f,g](L^n)=(G_{34}[\tilde{g}]\circ
G_{12}[f])(L^n)=(\sigma L^n \sigma^{-1})_{\geq 1}+\sum_{j=1}^N\hat{q}_j\partial^{-1}\hat{r}_j\p\\
&=&\p^n+\hat{\pi}_{n-1}\p^{n-1}+\cdots+\hat{\pi}_1\p+\sum_{j=1}^N\hat{q}_j\partial^{-1}\hat{r}_j\p,\\
&& \sigma=\p^{-1}\tilde{g}_x\tilde{g}^{-2}\p^{-1}\tilde{g}\p
f_x^{-1}f^2\p
f^{-1},\ \ (\tilde{g}=f^{-1}D^{-1}(gf_x)),\\
\end{array}
\end{equation}
\begin{equation}
\label{242}
\hat{\pi}_{n-1}=n\hat{v},\ \ \hat{v}=v-[{\mathrm{ln}}(\frac{D^{-1}(fg_x)}{D^{-1}(f_xg)})]_x,\\
\end{equation}
\begin{equation}
\label{243}
\hat{q}_j\stackrel{\triangle}{=}G[f,g](q_j)=(G_{34}[\tilde{g}]\circ
G_{12}[f])(q_j)=q_j-\frac{fD^{-1}(gq_{j,x})}{D^{-1}(gf_x)}, \\
\end{equation}
\begin{equation}
\label{244}
\hat{r}_j\stackrel{\triangle}{=}G[f,g](r_j)=(G_{34}[\tilde{g}]\circ
G_{12}[f])(r_j)=r_j-\frac{gD^{-1}(fr_{j,x})}{D^{-1}(fg_x)},\  \
j=1,...,N.\\
\end{equation}
\end{subequations}

{\bf{Remark}}:\\

1.When $q_j=r_j=0$,$j=1,\cdots,N$, $G_i$,$G_{kl}$ and $G$
introduced above will degenerate to the auto-B\"{a}cklund
transformations for the mKP hierarchy based on gauge
transformations which have been widely studied in [24-31](Now $f$
only need to be an eigenfunction of the mKP hierarchy, i.e.,
satisfy (\ref{9}) with $q_i=r_i=0$. $g$ only need to be an adjoint
eigenfunction (in some articles, often called integrated adjoint
eigenfuction) of the mKP hierarchy,
i.e., satisfy (\ref{10}) with $q_i=r_i=0$);\\

2.When $k=2$ and $N=3$, $G[f,g]$ introduced above will give rise
to the auto-B\"{a}cklund transformation for the mKPESCS
(\ref{10.1}) induced by binary Darboux transformation introduced
in \cite{Estevez} ($f$ and $g$ now only need to be eigenfunction
and adjoint eigenfunction of the mKPESCS
respectively).\\

3.({\bf{Iteration of $G$}}). \\Assuming $\phi_1,...,\phi_n$ are $n$
arbitrary solutions of (\ref{9}) and $\bar{\phi}_1,...,\bar{\phi}_n$
are $n$ arbitrary solutions of (\ref{10}), we define the following
Wronskians (Gramm Determinants):
\begin{equation}
\label{25}
\begin{array}{lll}
W_1(\phi_1,...,\phi_n;\bar{\phi}_1,...,\bar{\phi}_n)&=&det(Y_{n\times
n}),\\W_2(\phi_1,...,\phi_n;\bar{\phi}_1,...,\bar{\phi}_n)&=&det(\tilde{Y}_{n\times
n}),\\W_3(\phi_1,...,\phi_n;\bar{\phi}_1,...,\bar{\phi}_{n-1})&=&det(Z_{n\times
n}),\\W_4(\phi_1,...,\phi_{n-1};\bar{\phi}_1,...,\bar{\phi}_n)&=&det(\tilde{Z}_{n\times
n}),
\end{array}
\end{equation}
where
\begin{subequations}
\label{25.1}
\begin{equation}
\label{25.11}
     Y_{i,j}=D^{-1}(\bar{\phi}_j\phi_{i,x}),\ \ \  i,j=1,...,n,
\end{equation}
\begin{equation}
\label{25.12} \tilde{Y}_{i,j}=D^{-1}(\bar{\phi}_{j,x}\phi_i),\ \ \
i,j=1,...,n,
\end{equation}
\begin{equation}
\label{25.13} Z_{i,j}=D^{-1}(\bar{\phi}_i\phi_{j,x}),\
i=1,...,n-1,\ j=1,...,n;\ \ Z_{n,j}=\phi_j,\ j=1,...,n,
\end{equation}
\begin{equation}
\label{25.14} \tilde{Z}_{i,j}=D^{-1}(\bar{\phi}_{j,x} \phi_i),\
i=1,...,n-1,\ j=1,...,n;\ \ \tilde{Z}_{n,j}=\bar{\phi}_j,\ \
j=1,...,n.
\end{equation}
\end{subequations}
We denote $v[n]$, $q_i[n]$ and $r_i[n]$ as the $n$ times iteration
of $G$ on $v$, $q_i$ and $r_i$ respectively, then the following
formulas hold.
\begin{subequations}
\label{25.2}
\begin{equation}
\label{25.21}
     v[n]=v-\partial_x{\mathrm{ln}} \frac{W_2(f_1,...,
f_n;g_1,..., g_n)}{W_1(f_1,..., f_n;g_1,..., g_n)},
\end{equation}
\begin{equation}
\label{25.22} q_i[n]=\frac {W_3(f_1,..., f_n,q_i;g_1,...,
g_n)}{W_1(f_1,..., f_n;g_1,..., g_n)},
\end{equation}
\begin{equation}
\label{25.23} r_i[n]=\frac {W_4(f_1,..., f_n;g_1,...,
g_n,r_i)}{W_2(f_1,..., f_n;g_1,..., g_n)},
\end{equation}
\end{subequations}
where $f_i$ and $g_i$, $i=1,\cdots,n$ are arbitrary solutions of
(\ref{9}) and (\ref{10}) respectively. \\
We can prove (\ref{25.2}) in the same way as we did in
\cite{XiaoTing20042} and we omit it here.

\section{The B\"{a}cklund transformations between the KPHSCS and mKPHSCS}
\setcounter{equation}{0} \hskip\parindent

(1) B\"{a}cklund transformation utilizing eigenfunction of the
KPHSCS.\\
\begin{theorem}
Let $L^n$ of (\ref{2}) satisfy the KPHSCS (\ref{3}) and $\psi$ be
the corresponding eigenfunction. The function $f\neq 0$ satisfies
(\ref{4}). Define
$$M_1[f]: L^n \stackrel{f}{\longmapsto}
\tilde{L}^n$$ by
\begin{equation}
\label{26} \tilde{L}^n=(f^{-1}L^nf)_{\geq
1}+\sum_{j=1}^N\tilde{q}_j\partial^{-1}\tilde{r}_j\p,\ \
\tilde{q}_j=f^{-1}q_j, \ \
\tilde{r}_j=-D^{-1}(r_jf),\ \ j=1,...,N.\\
\end{equation}
Then $\tilde{L}^n$ will satisfy the mKPHSCS (\ref{8}).
\end{theorem}
{\bf{Proof}}:It is straightforward to see that
$$(\tilde{L}^n)_{\geq 1}=(f^{-1}L^nf)_{\geq 1}=\p^n+\tilde{\pi}_{n-1}\p^{n-1}+\cdots+\tilde{\pi}_1\p,\ \
\tilde{\pi}_{n-1}=n\tilde{v}=n(({\mathrm{ln}} f)_x).$$ Furthermore
we can find that for $k<n$,
\begin{equation}
\nonumber
\begin{array}{lll}
(\tilde{L}^k)_{\geq 1}&=&[(\tilde{L}^n)^{\frac{k}{n}}]_{\geq
1}=[((\tilde{L}^n)_{\geq 1})^{\frac{k}{n}}]_{\geq 1}=[((f^{-1}
L^nf)_{\geq 1})^{\frac{k}{n}}]_{\geq
1}=[(f^{-1} L^nf)^{\frac{k}{n}}]_{\geq 1}\\
&=&[f^{-1} (L^n)^{\frac{k}{n}}f]_{\geq 1}=[f^{-1} (L^k)_{\geq
0}f]_{\geq 1}.
\end{array}
\end{equation}

Define $\tilde{\phi}=f^{-1}\psi$. From Lemma3.2 1), we can easily
see that $\tilde{q}_i$,$\tilde{r}_i$,$i=1,...,N$ and
$\tilde{\phi}$ satisfy (\ref{82}),(\ref{83}) and (\ref{91}) w.r.t.
$\tilde{L}^n$ respectively. Besides,
\begin{equation}
\nonumber
\begin{array}{lll}
\tilde{\phi}_{t_n}-(\tilde{L}^n)_{\geq 1}(\tilde{\phi})&=&-f^{-2}\psi[f_{t_n}-(L^n)_{\geq 0}(f)]+f^{-1}[\psi_{t_n}-(L^n)_{\geq 0}(\psi)]\\
&=&-f^{-2}\psi\s q_iD^{-1}(r_if)+f^{-1}[\s q_iD^{-1}(r_i\psi)]\\
&=&-\s f^{-1}q_i[\tilde{\phi}D^{-1}(r_if)-D^{-1}(r_if\tilde{\phi})]\\
&=&\s \tilde{q}_iD^{-1}(\tilde{r}_i\tilde{\phi}_x),
\end{array}
\end{equation}
so $\tilde{\phi}$ satisfies (\ref{92}) w.r.t.
$\tilde{L}^n$.\\
This completes the proof.\\

(2) B\"{a}cklund transformation utilizing adjoint eigenfunction of
the KPHSCS.\\
\begin{theorem}
Let $L^n$ of (\ref{2}) satisfy the KPHSCS (\ref{3}) and $\psi$ be
the corresponding eigenfunction. The function $g\neq 0$ satisfies
(\ref{5}). Define
$$M_2[g]: L^n \stackrel{g}{\longmapsto}
\tilde{L}^n$$ by
\begin{equation}
\label{28} \tilde{L}^n=(\p^{-1}gL^ng^{-1}\p)_{\geq
1}+\sum_{j=1}^N\tilde{q}_j\partial^{-1}\tilde{r}_j\p, \ \
\tilde{q}_j=D^{-1}(gq_j),\ \ \tilde{r}_j=r_jg^{-1},\ \
j=1,...,N.\\
\end{equation}
Then $\tilde{L}^n$ will satisfy the mKPHSCS (\ref{8}).
\end{theorem}
{\bf{Proof}}:It is straightforward to see that
$$(\tilde{L}^n)_{\geq 1}=(\p^{-1}gL^ng^{-1}\p)_{\geq 1}=\p^n+\tilde{\pi}_{n-1}\p^{n-1}+\cdots+\tilde{\pi}_1\p,\ \
\tilde{\pi}_{n-1}=n\tilde{v}=-n(({\mathrm{ln}} g)_x).$$
Furthermore we can find that for $k<n$,
\begin{equation}
\nonumber
\begin{array}{lll}
(\tilde{L}^k)_{\geq 1}&=&[(\tilde{L}^n)^{\frac{k}{n}}]_{\geq
1}=[((\tilde{L}^n)_{\geq 1})^{\frac{k}{n}}]_{\geq 1}=[((\p^{-1}g
L^ng^{-1}\p)_{\geq 1})^{\frac{k}{n}}]_{\geq
1}=[(\p^{-1}g L^ng^{-1}\p)^{\frac{k}{n}}]_{\geq 1}\\
&=&[\p^{-1}g (L^n)^{\frac{k}{n}}g^{-1}\p]_{\geq
1}=[\p^{-1}g(L^k)_{\geq 0}g^{-1}\p]_{\geq 1}.
\end{array}
\end{equation}
Define $\tilde{\phi}=D^{-1}(g\psi)$. From Lemma3.3 1), we can
easily see that $\tilde{q}_i$,$\tilde{r}_i$, $i=1,...,N$, and
$\tilde{\phi}$ satisfy (\ref{82}),(\ref{83}) and (\ref{91}) w.r.t.
$\tilde{L}^n$ respectively. Besides,
\begin{equation}
\nonumber
\begin{array}{lll}
\tilde{\phi}_{t_n}-(\tilde{L}^n)_{\geq 1}(\tilde{\phi})&=&D^{-1}[\psi(g_{t_n}+(L^n)_{\geq 0}^*(g))]+D^{-1}[g(\psi_{t_n}-(L^n)_{\geq 0}(\psi))]\\
&=&D^{-1}[\psi\s r_iD^{-1}(q_ig)]+D^{-1}[g\s q_iD^{-1}(r_i \psi)]\\
&=&\s D^{-1}(gq_i)D^{-1}(r_i\psi)\\
&=&\s \tilde{q}_iD^{-1}(\tilde{r}_i\tilde{\phi}_x).\\
\end{array}
\end{equation}
so $\tilde{\phi}$ satisfies (\ref{92}) w.r.t.
$\tilde{L}^n$.\\
This completes the proof.\\

\section*{ Acknowledgment}\hskip\parindent
This work was supported by the Chinese Basic Research Project
"Nonlinear Science".

\hskip\parindent

\end{document}